\documentstyle[12pt]{article}
\newcommand{\BEQ}{\begin{equation}}    
\newcommand{\BEA}{\begin{eqnarray}}
\newcommand{\EEQ}{\end{equation}}      
\newcommand{\EEA}{\end{eqnarray}}
\newcommand{\eps}{\epsilon}                      
\newcommand{\sig}{\sigma}                        
\newcommand{\rar}{\rightarrow}                   
\newcommand{\zeile}[1]{\vskip #1 \baselineskip}  

\newcommand{\build}[3]{\mathrel{\mathop{\kern 0pt#1}\limits_{#2}^{#3} }}

\catcode`\@=11
\def\numberbysection{\@addtoreset{equation}{section}
        \def\theequation{\thesection.\arabic{equation}}}

\begin{document}
\baselineskip 0.3in
%
%
\begin{titlepage}
\null
\begin{center}
\vskip 0.5in
{\Large \bf Boundary-Induced Phase Transitions in Equilibrium and
Non-Equilibrium Systems}
\vskip 0.5in
Malte Henkel$^{a,{\S}}$ and Gunter Sch\"utz$^{b,{\S}}$
 \\[.3in]
{\em $^{a}$D\'epartement de Physique Th\'{e}orique,
     Universit\'e de Gen\`eve \\
     24  quai Ernest Ansermet,
     CH - 1211 Gen\`eve 4, Switzerland}
\zeile{1}
{\em $^{b}$Department of Physics, Weizmann Institute of Science, \\
Rehovot 76100, Israel}
\zeile{2}
{\bf UGVA-DPT 1993/07-826} \\
{\bf WIS-93/73/Aug.-PH}
\end{center}
\zeile{2}
%
\begin{abstract}
Boundary conditions may change the phase diagram of non-equilibrium
statistical systems like the one-dimensional
asymmetric simple exclusion process with and without particle number
conservation.
Using the quantum Hamiltonian approach, the model is mapped onto an XXZ
quantum chain and solved using the Bethe ansatz.
This system is related to a two-dimensional vertex model in thermal
equilibrium. The phase transition caused by a point-like boundary
defect in the dynamics of the one-dimensional exclusion model is in
the same universality class as a
continous (bulk) phase transition of the
two-dimensional vertex model caused by a line defect at its boundary.
\end{abstract}
\zeile{1}
$^{\S}$ Address after 1 October 1993: {\em Department of Theoretical
Physics, University of Oxford, 1 Keble Road, Oxford OX1 3NP, UK}

\end{titlepage}

\newpage
%
%

It is generally held that for statistical systems the influence of
boundary conditions should be negligible, at least when one is
interested in the bulk system only. This belief is indeed supported
for equilibrium systems by abundant evidence coming from analytical
or numerical studies. As we are going to show in this letter, however,
this picture is far from being generally valid. Boundary conditions
may indeed be crucial for the bulk properties and may even cause
continous phase transitions.

It has already been realized that a change
in boundary conditions (equivalent to some localized point defect
in a system on a ring) may cause various kinds of phase transitions
in the {\em static} properties of
one-dimensional {\em non-equilibrium} systems \cite{1,2}.
Here we show that boundary terms may also induce phase
transitions in the {\em dynamics} of these systems.
Such phase transitions are then shown to correspond to bulk
phase transitions of two-dimensional {\em equilibrium} systems caused
by boundary terms (i.e., line defects).

The paradigmatic example we are going to study is the asymmetric
simple exclusion process, see \cite{Ligg85}.
In its simplest version, without particle creation or annihilation,
it is described by particles of a single species $A$ moving on
a lattice. A given site $j$ can be occupied
or empty at an instant of time $t$. A particle at site
$j$ for time $t$ may hop at time $t+1$ to its right neighbor
with rate $(1+\eps)/2$ and to its left neighbor with rate $(1-\eps)/2$,
if the final site is empty. This simple model appears in a large variety
of contexts. It has been argued to be in the same universality class
as the noisy Burger's equation \cite{Gwa92}.
This in turn is a one-dimensional
version of the incompressible Navier-Stokes equation or
else can be regarded
as the one-dimensional Kardar-Zhang-Parisi equation describing the
shape fluctuation in various growth models.

Using a master equation approach,
the probability distribution function $P(\{\beta\};t)$ is obtained
by solving $\partial_{t} P = -H P$, where $\beta(t)$
is a configuration of occupied and empty sites and the
quantum Hamiltonian reads \cite{Gwa92}
\BEQ \label{eq:GS}
H = -\frac{1}{4} \sum_{j=1}^{L} \left[
\vec{\sig}_j \cdot \vec{\sig}_{j+1} -1
+i\eps \left( \sig_{j}^{x}\sig_{j+1}^{y}-\sig_{j}^{y}\sig_{j+1}^{x}
\right)\right]
\EEQ
where $L$ is the number of sites and $\sig^{x,y,z}$ are Pauli matrices.
The ground state energy vanishes as
follows from probability conservation in the master equation.
Obviously $H$ commutes with the particle number operator
$N= \sum_{j=1}^{L} n_j$ where we have introduced the particle projector
$n_j = \frac{1}{2} (1-\sig_{j}^{z})$. The particle density
is then $\rho = N/L$. Using periodic boundary conditions
$\sig_{L+1}^{x,y,z}=\sig_{1}^{x,y,z}$,
Gwa and Spohn \cite{Gwa92} then show using
Bethe ansatz techniques for $\eps=1$ and $\rho=1/2$ that the
(real part of the) energies $E_L$ for $L$ large scale as
\BEQ \label{eq:GSPeri}
E_L \sim L^{-3/2}
\EEQ

On the other hand, the Hamiltonian had also been derived considering
asymmetric lattice diffusion for the case
of free boundary conditions where the particles are not allowed
to move beyond the boundary (impenetrable walls).
Then it reads \cite{Alca93}
\BEQ \label{eq:ADHR}
H' = - \frac{1}{4\Delta}
\sum_{j=1}^{L-1} \left[ \sig_{j}^{x} \sig_{j+1}^{x}
+ \sig_{j}^{y} \sig_{j+1}^{y} + \Delta \sig_{j}^{z} \sig_{j+1}^{z}
 - \frac{q - q^{-1}}{2} \left(
\sig_{j}^{z} - \sig_{j+1}^{z} \right)  - \Delta \right]
\EEQ
where
\BEQ \label{eq:DelQ}
\Delta = \frac{q+q^{-1}}{2} \;\; , \;\; q = \sqrt{\frac{1-\eps}{1+\eps}}
\EEQ
This is the well-known XXZ quantum chain which is symmetric under
the quantum group $U_{q}SU(2)$ as shown
by Pasquier and Saleur \cite{Pasq90}.
We stress that the quantum Hamiltonian of
the diffusion process on a periodic
lattice cannot be obtained by simply taking
periodic boundary conditions in
eq.~(\ref{eq:ADHR}).
Using the $U_qSU(2)$ symmetry it is easy to show
that for $\eps \neq 0$
the energies for $L$ large become
\BEQ \label{eq:ADHRFrei}
{E_L}' \sim  1 - \Delta^{-1}
\EEQ
in each sector with $N$ particles, see \cite{Pasq90,McCoy,SS}.

The apparent inconsistency of
eqs.~(\ref{eq:GSPeri},\ref{eq:ADHRFrei}) will
be explained below. Our analysis applies to any value of $\eps$,
generalising the approach of \cite{Gwa92}.
We note that since the Hamiltonian of
eq.~(\ref{eq:GS}) is gapless, the approach of time-dependent
mean values towards their equilibrium value may be according to
a power law, while with the Hamiltonian of
eq.~(\ref{eq:ADHR}) the relaxation will be always exponential.

To understand this observation, we begin by rewriting
eq.~(\ref{eq:GS}).
Introducing raising and lowering operators
$\sig^{\pm}=\frac{1}{2}(\sig^{x}\pm i \sig^{y})$, we have
\BEA
H &=& -\frac{1+\eps}{2} \sum_{j=1}^{L} \left[ \frac{1-\eps}{1+\eps}
\sig_{j}^{+}\sig_{j+1}^{-} + \sig_{j}^{-}\sig_{j+1}^{+}
+\frac{1}{2(1+\eps)} \left(
\sig_{j}^{z}\sig_{j+1}^{z} -1\right) \right]
\nonumber \\
&=& -\frac{1}{q+q^{-1}} \sum_{j=1}^{L} \left[ q
\sig_{j}^{+}\sig_{j+1}^{-} + q^{-1} \sig_{j}^{-}\sig_{j+1}^{+}
+\frac{q+q^{-1}}{4} \left( \sig_{j}^{z}\sig_{j+1}^{z} -1\right) \right]
\EEA
where the second of the eqs.~(\ref{eq:DelQ}) has been used. Next,
consider the non-singular operator
\BEQ \label{eq:U}
U = \exp \left( \pi g \sum_{j=1}^{L} j \sig_{j}^{z} \right) \;\; , \;\;
U \sig_{j}^{\pm} U^{-1} = e^{\pm 2\pi g j} \sig_{j}^{\pm}
\EEQ
If we now choose $q=e^{2\pi g}$, we obtain
\BEQ \label{eq:H}
H'' = U H U^{-1} = - \frac{1}{2(q+q^{-1})}
\sum_{j=1}^{L} \left[ \sig_{j}^{x} \sig_{j+1}^{x}
+ \sig_{j}^{y} \sig_{j+1}^{y} + \Delta \sig_{j}^{z} \sig_{j+1}^{z}
- \Delta \right]
\EEQ
which is indeed (almost) the Hamiltonian $H'$.
The distinction comes from the boundary conditions.
The surface field $(q-q^{-1})(\sigma_1^z-\sigma_L^z)/8\Delta$
is absent in $H''$ and one has
\BEQ \label{eq:BC}
\sig_{L+1}^{\pm} = q^{\mp L} \sig_{1}^{\pm} \;\; , \;\;
\sig_{L+1}^{z} = \sig_{1}^{z}
\EEQ
which make $H''$ non-hermitian, as $H$ already is.
It is these unusal boundary conditions which give rise to the
different properties of $H$ and $H'$, as we now show.

First we note that the ground state of the system (\ref{eq:ADHR})
with free boundary conditions as well as that of the model (\ref{eq:H})
with twisted boundary conditions (\ref{eq:BC})
is $(L+1)$-times degenerate and has
energy 0, independent of $L$. In what follows, we assume $L$ to
be even. In each sector with fixed particle
number $N$ the lowest energy is 0. This can be shown by either
solving the Bethe ansatz equations or by using the symmetry relations
with the generators of the quantum algebra \cite{Pasq90,SS,GS}.

The calculation of the spectrum proceeds via the Bethe ansatz. Indeed,
the XXZ chain with non-periodic boundary conditions was already
considered in \cite{Alca88} and we merely have to adapt their
results to the problem at hand. To begin with, we consider the
sector with $N=1$ particle. Then the energies are in this sector
\BEQ\label{eq:energy}
E = 1- \Delta^{-1} \cos \theta \;\; , \;\;
\theta = 2\pi \left(i g + \frac{n}{L}\right)
\EEQ
where $n$ is an integer from the set
$\{ 0, \pm 1,\ldots, \pm(L/2 -1),L/2\}$. For $L$ large, the
energies become
\BEA\label{eq:gap}
E &=& 1 - \frac{\cos\left( 2\pi \left(i g +\frac{n}{L}\right)\right)}
{\cosh (2\pi g)} \nonumber \\
&\simeq & 2\pi^2 \left(\frac{n}{L}\right)^{2}
+ 2\pi i \tanh(2\pi g) \frac{n}{L} + \ldots
\EEA
and we note that the mass term indeed cancels. We also observe that the
real part $\Re E \sim L^{-2}$ and the imaginary part
$\Im E \sim L^{-1}$. We return to an explanation of this below.

Next, we take the sector $N=2$. From the Bethe ansatz \cite{Alca88}
we have
\BEA
&& E = (2 \Delta - cos \theta -\cos \theta')\Delta^{-1} \\
&& \theta L -2\pi i g L= 2\pi I - \Theta(\theta,\theta') \;\; , \;\;
\theta' L -2\pi i g L= 2\pi I' - \Theta(\theta',\theta) \nonumber
\EEA
where $I,I'$ are distinct half-integers from the set
$\{\pm\frac{1}{2},\pm\frac{3}{2},\ldots,\pm\frac{L-1}{2}\}$ and
\BEQ
\Theta(\theta,\theta') = 2 \arctan
\left( \frac{\Delta \sin\left( \frac{1}{2}
(\theta-\theta')\right)}{\cos\left(\frac{1}{2}(\theta+\theta')\right)
-\Delta \cos\left(\frac{1}{2}(\theta-\theta')\right)} \right)
\EEQ
Define $\widetilde{\theta}=\theta-2\pi i g$,
$\widetilde{\theta'}=\theta'-2\pi i g$. Then, for small values of
the arguments
\BEQ
\Theta(\widetilde{\theta},\widetilde{\theta'}) \simeq
2 \arctan\left( i \coth(2\pi g) \cdot
\frac{\widetilde{\theta}-\widetilde{\theta'}}{
\widetilde{\theta}+\widetilde{\theta'}} \right)
\EEQ
which is of order unity. Therefore
\BEQ
\widetilde{\theta} = \frac{2\pi}{L} a \;\; , \;\;
\widetilde{\theta'} = \frac{2\pi}{L} a'
\EEQ
where $a,a'$ are of order ${\cal O}(1)$.
It follows that the same cancellation
as observed in the sector $N=1$ also takes place
here and also that the observed
scaling of the energies does not change. Finally, for $N$ arbitrary
\BEQ\label{eq:spectrum}
E = \Delta^{-1} \left( N \Delta - \sum_{n=1}^{N} \cos \theta_n \right)
\EEQ
with
\BEQ
\theta_m L -2\pi i g L=
2\pi I_m - \sum_{n=1}^{N} \Theta(\theta_m,\theta_n)
\EEQ
and the same argument can be repeated. It follows that for any value
of $\eps$, the spectrum is massless. There is a distinction, however,
between finite particle densities $\rho = N/L = {\cal O}(1)$ and
small densities $\rho = {\cal O}(L^{-1})$. As shown above,
in the latter case the
real part of the energy gaps scales as $L^{-2}$, while for $\rho=1/2$
it vanishes like $L^{-3/2}$ (when $\eps=1$) \cite{Gwa92}.
The way how this result
was achieved suggests that this scaling behaviour is characteristic
for all finite densities.

After this analysis of the spectra of $H'$ and $H''$ (or $H$, as
the spectra of $H$ and $H''$ are identical) we turn to an interpretation
of our results. The $N$-particle ground state $|N\rangle$
is stationary with respect to the stochastic
process defined by the Hamiltonian and we shall refer to it as
the steady state of the system. Average values $\langle C \rangle$
of operators $C$ in some $N$-particle probability
distribution $| P_N \rangle$
are defined as $\langle C \rangle = \langle N | C | P_N \rangle$.
Of particular interest
are the $n$-point  correlation functions in the steady state
\BEQ\label{eq:corr}
G(x_n,t_n; \dots ; x_1,t_1; 0,0) =
\langle N | n_{x_n} \mbox{e}^{-H(t_n - t_{n-1})} \dots
n_{x_1} \mbox{e}^{-H t_1 } n_0 | N \rangle
\EEQ
and their connected counterparts $G_c$ defined by replacing the
operators $n_x$ by $(n_x - \rho)$ in (\ref{eq:corr}).

We discuss the system dynamics by studying the dynamic structure
function $S_N(k,t)$ which is the Fourier transform of the connected
two-point correlation function in the $N$-particle sector.
In the case of asymmetric diffusion with free boundary conditions
(\ref{eq:ADHR}) the correlation function decays exponentially
in time with finite relaxation time $\tau = \Delta/(\Delta -1)$,
see eq.~(\ref{eq:ADHRFrei}).
\BEQ\label{eq:S1}
S_N(k,t) \propto f(k) \mbox{e}^{-t/\tau}
\EEQ
The relaxation time does not depend
on the density $\rho$. In the first model, however, corresponding
to asymmetric diffusion with periodic boundary conditions,
the relaxation time diverges. For small densities (finite number of
particles) we read from eq.~(\ref{eq:gap})
$\tau \sim L^2$ while the result obtained by
Gwa and Spohn translates into $\tau \sim L^{3/2}$ for finite
densities.
For the dynamic structure function this suggests the
scaling form \cite{Gwa92}
\BEQ\label{eq:S2}
S(k,t) \propto \mbox{e}^{-i\epsilon (1-2\rho) k t} h( k^{3/2} t)
\EEQ
for finite densities while computing the exact structure function
for one particle from eq.~(\ref{eq:energy}) suggests
\BEQ\label{eq:S3}
S(k,t) \propto \mbox{e}^{-i\epsilon k t}
\mbox{e}^{-\frac{1}{2}k^2 t}
\EEQ
for any finite number of particles. The phase factors $\exp{(-i
\epsilon (1-2\rho) t)}$ arise from the imaginary part of the energy
gaps and reflect the steady state current of
particles moving around the ring (recall $\tanh (2\pi g)=-\eps$).
The scaling functions $h(k^{3/2}t)$
and $\exp{(-k^2 t/2)}$ have their origin in the diffusive nature of the
process. From these expressions one can read the dynamic exponent of the
process: it is $z=2$ for small densities and $z=3/2$ for finite
densities. It is interesting to note the consistency with the
results for the diffusion constant $D$ of a
single tagged particle in the
fully asymmetric exclusion model studied by Derrida et {\em al.}
\cite{DEM}. These authors consider an exclusion model of the kind
discussed here containing one particle which has the same dynamics
as all the other (indistinguishable) particles, but is tagged.
(Effectively, this a model with two types of particles, A and
B, containing $N_A$ particles of type A and $N_B=1$ particle of
type B all moving with the same probability to the right if the
neighbouring site was empty.) They follow the motion of the tagged
particle and compute its diffusion constant $D$.
They find  that $D$ is finite for a finite number of untagged
particles, corresponding to an dynamical exponent $z=2$.
On the other hand $D$ diverges proportional to $L^{1/2}$
for finite densities which corresponds to $z=3/2$.

The discussion of behaviour of the dynamic structure function
for free boundary conditions (\ref{eq:S1}) on the one hand
and for twisted (periodic) boundary conditions
(\ref{eq:S2}) and (\ref{eq:S3}) on the other hand
shows that a phase transition takes place when
changing the boundary conditions of the model and elucidates its
effect on the dynamics of the system. Now we show that this
corresponds to a bulk phase transition of a
two-dimensional six-vertex model. The
Hamiltonians $H_L'$ and $H_L''$ of the model with $L$ sites
can be derived from the transfer matrix
$T_L$ of the six-vertex model with a defect line of the type
given in eq.~(\ref{eq:BC}) located between
sites $L$ and 1 (which we call the boundary) \cite{2,Gwa92}.
Its partition function $Z$ in thermal
equilibrium on a $L \times M$ lattice is given by the trace of the
$M^{th}$ power of $T_L$. In the thermodynamic limit $L,M
\rightarrow \infty$ one obtains $Z = \lim_{L,M \rightarrow \infty}
\mbox{Tr} \left[  \exp{(-H_L M)} \right]$.
{}From the vanishing of the energy gaps discussed above
(which corresponds to the appearance of an
infinite degeneracy of the ground state of $H$)
we conclude that changing the boundary conditions causes a
continous phase transition in the two-dimensional six-vertex model in
thermal equilibrium.

So far we have studied asymmetric diffusion with a conserved number
of particles. We briefly show that a boundary induced phase transition
occurs also in an asymmetric diffusion model with
pair annihilation $A+A\rar\emptyset$. We study the Hamiltonian
\BEQ\label{eq:H2}
H = - \frac{1}{2(q+q^{-1})}
\left(\sum_{j=1}^{L-1} h_j + b \; h_L \right)
\EEQ
with
\BEQ\label{eq:u}
h_j = q \sig_{j}^{+}\sig_{j+1}^{-} + q^{-1} \sig_{j}^{-}\sig_{j+1}^{+}
+\frac{q+q^{-1}}{2}  \left(\sig_{j}^{+}\sig_{j+1}^{+} -2 \right)
+ q \sig_j^z + q^{-1} \sig_{j+1}^z \hspace*{2mm} .
\EEQ
The boundary conditions are defined by the parameter $b$.
If  $b=0$ one has free boundary conditions (we call this Hamiltonian
$H^F$ when a distinction w.r.t. the boundary conditions is necessary)
while $b=1$ correspond to periodic
boundary conditions (denoted $H^P$).
$H$ describes a process where particles in a pair
of sites hop with probability $(1+\epsilon)/2$ to the right and
probability $(1-\epsilon)/2$ to the left if the respective sites
are empty. Pairs of particles are always annihilated.
$H$ has a twofold degenerate steady state
with energy 0.
These are the ferromagnetic ground
 state containing no particles (all spins up)
and the one-particle state (one spin down)
in which each possible position of the particle is equally probable.

The operator $U$ eq.~(\ref{eq:U}) transforms $H^F$ into the
free fermion Hamiltonian studied in \cite{Alca93} which describes a
transition of the Pokrovsky-Talapov type \cite{Pokr80}. The energy
gaps  are all finite and of the form
\BEQ
E^F \sim N \left( 1- \frac{2}{q+q^{-1}} \right)
\EEQ
for large $L$, which implies an exponentially slow approach to the
stationary state at late times.

For periodic boundary conditions $H^P$ is transformed by $U$ into
\BEQ
{H^P}' = - \frac{1}{2(q+q^{-1})} \sum_{j=1}^{L}
\left[ \sig_{j}^{x} \sig_{j+1}^{x} + \sig_{j}^{y} \sig_{j+1}^{y}
+ \frac{q+q^{-1}}{2} \left( q^{2j+1} \sig_{j}^{+} \sig_{j+1}^{+}
- 2 \right) + q\sig_{j}^{z}+q^{-1}\sig_{j+1}^{z} \right]
\EEQ
with the twisted boundary conditions eq.~(\ref{eq:BC}). Following the
argument of \cite{Alca93}, the spectrum of ${H^P}'$ is seen to be
independent of the annihilation term
$\sig_j^{+}\sig_{j+1}^{+}$ and it is therefore equal to the spectrum of
\BEQ
{H_0^P}' =  - \frac{1}{2(q+q^{-1})} \sum_{j=1}^{L}
\left[ \sig_{j}^{x} \sig_{j+1}^{x} + \sig_{j}^{y} \sig_{j+1}^{y}
+ \frac{q+q^{-1}}{2} \left( \sig_{j}^{z}+\sig_{j+1}^{z} - 2 \right)
\right]
\EEQ
which commutes with the particle number operator $N$ and
describes non-interacting fermions on a ring with twisted
boundary conditions eq.~(\ref{eq:BC}). The eigenvalues of ${H_0^P}'$
are easily found
by a Jordan-Wigner and Fourier transformation and
are of the form
\BEQ
E^P = \sum_{i=1}^{N} \left( 1- \frac{2}{q+q^{-1}} \cos{\theta_i} \right)
\EEQ
similar to eq.~(\ref{eq:spectrum}) but with
\BEQ
\theta_i = 2\pi \left( ig + m_i/L \right)
\EEQ
where the $m_i$ are pairwise distinct integers $0 \leq m_i \leq L-1$ for
$N$ odd and half-integers $\frac{1}{2} \leq m_i \leq L-\frac{1}{2}$ for
$N$ even. The low lying energy gaps ($N$ finite) vanish in the
thermodynamic limit
as $\Re E^P \sim L^{-2}$ and $\Im E^P \sim L^{-1}$.
Here, as opposed to the model with free boundary conditions,
finite-density states decay with a relaxation time of order 1, while for
low-density states (finite $N$) the relaxation time diverges
proportional to $L^2$.

The Hamiltonian (\ref{eq:H2})
is related to a 7-vertex model with a boundary defect,
see \cite{Alca93}. As in the
6-vertex model discussed above, changing the boundary condition
induces a bulk phase transition. The same mechanism should also
apply to more general reaction-diffusion problems,
for example two-particle processes with a reaction
$A+B\rar\emptyset$ \cite{Priv92}, or even extensions of the Hubbard
model \cite{Alca93a}, where the
same quantum chains as considered here reappear \cite{Alca93}.

To summarize, we have shown that boundary conditions can have a major
influence on the phase diagram of certain statistical models.
Since the models considered here are merely prototypes of much more
general ones, we expect that the phenomenon found is generic.

\zeile{2}
\noindent{\bf Acknowledgements}
\zeile{1.5}
\noindent M.H. thanks the Swiss National Science Foundation for
support. G.S. would like to thank H. Spohn for useful discussions.
Financial support by the Deutsche Forschungsgemeinschaft is gratefully
acknowledged.

\end{document}